\documentstyle [12pt]{article}

\newcommand{\ket}[1]{\left|\,#1\right\rangle}

\newcommand{\bk}[2]{\left\langle #1,#2 \right\rangle}
\begin{document}
\begin{titlepage}
\begin{flushleft}
\noindent ITP 92-09\\
April 1992
\end{flushleft}
\vspace{1.0 cm}
\begin{center}
{\huge {\bf Localization of Massless Spinning}} \vspace{0.5cm}\\
{\huge {\bf Particles and
the Berry Phase}}\\
\vspace{2.0 cm}
{\Large B.-S. Skagerstam}\footnote{Email
address: tfebss@secthf51.bitnet or tfebss@fy.chalmers.se. Research supported
by the Swedish National Research Council under contract no. 8244-103} \\
{\it Institute of Theoretical Physics \\
Chalmers University of Technology \\
S-412 96 G\"{o}teborg \\
Sweden}\\
\vspace{0.5cm}
{\Large Contribution to the Festschrift for John R. Klauder on occasion
 of his 60th birthday}
\end{center}
\vspace{0.5cm}
\begin{abstract}
The components of the position operator, at a fixed time, for a massless and
spinning
particle with given helicity $\lambda$ described in terms of bosonic degrees
of freedom have an anomalous commutator proportional to $\lambda$. The position
operator generates translations in momentum space. We show that a
ray-representation for these translations emerges due to the non-commuting
components of the position operators and relate this to the Berry-phase for
photons. The
Tomita-Chiao experiment then gives support for this relativistic and
 quantum mechanical description of photons in terms of non-commuting position
operators. \end{abstract}
\end{titlepage}
\newpage
\setcounter{page}{1}
\begin{center}
\section{\sc Introduction}
\end{center}

The concept of localization of relativistic elementary systems has a long and
intriguing history in physics (see e.g. Refs \cite{newton,wightman,jpa,local}).
Observations of physical phenomena takes place in space and time.
The notion of {\it localizability} of particles, elementary or not, then refers
to the
empirical fact that particles, at a given instance of time, appear to be
localizable in
the physical space.

In the realm of non-relativistic quantum mechanics the concept of
localizability of particles is built into the theory at a very
fundamental  level and is expressed in terms of the canonical commutation
relation
between a position operator and the corresponding generator of translations,
i.e. the canonical momentum of a particle. In relativistic theories the
concept of localizability of physical systems is  deeply connected to our
notion of space-time, the arena of physical phenomena, as a 4-dimensional
continuum. In the context of the general theory of relativity
 the localization of light rays in space-time is e.g. a fundamental
ingredient. In fact, it has been argued \cite{ehlers} that the Riemannian
metric is basically determined by basic properties of lightpropagation.

In a fundamental paper by Newton and Wigner \cite{newton} it
 was argued that in the context of relativistic quantum mechanics a notion of
point-like localization of a single particle
can be, uniquely, determined by kinematics. Wightman
\cite{wightman} extended this notion to localization to finite domains of
space and it was, rigourously, shown that massive particles are always
localizable if they are elementary,
i.e. if they are described in terms of
irreducible representations of the Poincar\'{e} group \cite{wigner}.
Massless
elementary systems with non-zero helicity, like a gluon, graviton,
neutrino or a photon, are not localizable in the sense of Wightman. The
axioms used by Wightman can, of course, be weakened. It was actually shown
by Jauch, Piron and Amrein \cite{jpa} that in such a sense the photon is
{\it weakly localizable}. The notion of weak localizability essentially
corresponds
to allowing for non-commuting observables in order to characterize the
localization of massless and spinning particles in general.

Localization of relativistic particles, at a fixed
 time, as alluded to above, has been shown to be
incompatible with a natural notion of (Einstein-) causality \cite{heger}.
If relativistic  elementary system has an exponentially small tail outside a
finite domain of localization at $t=0$, then, according to the hypothesis of
a weaker form of causality, this should remain true at later times, i.e. the
tail should only be shifted further out to infinity. As was shown by
Hegerfeldt \cite{heger2}, even this notion of causality is incompatible with
the notion of a positive and bounded observable whose expectation value gives
the probability to a find a particle inside a finite domain of space at a
given instant of time.

In the present paper we will reconsider some of these
questions related to the concept of localizability in terms of a quantum
mechanical description of a
massless particle with given helicity $\lambda $  \cite{baletal,atre,albs1}.
We will show how one can extend this description to include both positive and
negative
helicities. We will then be able to e.g. describe the motion of a linearly
polarized photon
in the framework of relativistic quantum mechanics.

\begin{center}
\section{\sc Position Operators for Massless Particles}
\end{center}
\setcounter{equation}{0}

It is easy to show that the components of the position operators for a massless
particle must
be non-commuting\footnote{This argument has, as far as we  know, first been
suggested by N. Mukunda.}  if the helicity $\lambda \neq 0$. If $J_{k}$ are the
generators
of rotations and $p_{k}$ the diagonal momentum, $k = 1,2,3$, then
 ${\bf J}\cdot {\bf p} = \pm \lambda$. Here ${\bf J } = (J_{1},J_{2},J_{3})$
and
${\bf p } = (p_{1},p_{2},p_{3})$ and ($\hbar = 1$)
\begin{equation}
[J_{k},p_{l}] = i\epsilon _{klm} p_{m}~~~~.
\end{equation}
If a canonical position operator ${\bf x }$ exists with components $x_{k}$
such that \begin{eqnarray}
&[x_{k},x_{l}] &= 0~~~~, \\
&[x_{k},p_{l}] &= i\delta _{kl}~~~~,
\label{ccr} \\
&[J_{k},x_{l}] &= i\epsilon _{klm} x_{m}~~~~,
\end{eqnarray}
then we can define generators of orbital angular momentum in the conventional
way, i.e.
\begin{equation}
L_{k} = \epsilon _{klm} x_{l}p_{m}~~~~.~
\end{equation}
 "Spin"- generators
are then defined by
\begin{equation}
 S_{k} = J_{k} - L_{k}~~~~.
\end{equation}
They fulfill the algebra
\begin{equation}
[S_{k}, S_{l}] = i\epsilon _{klm} S_{m}~~~~,
\end{equation}
and they, furthermore, commute with ${\bf x}$ and ${\bf p}$.
 Then, however, the spectrum
of ${\bf S}\cdot {\bf p}$ is $\lambda ,\lambda -1 ,..., -\lambda $, which
contradicts the requirement ${\bf J}\cdot {\bf p} = \pm \lambda$  since, by
construction,
${\bf J}\cdot {\bf p} ={\bf S}\cdot {\bf p}$.

 As has been discussed in detail
in the literature, the non-zero commutator of two components of the position
operator  for a
massless particle primarily emerge due to the non-trivial topology of the
momentum space \cite{baletal,atre,albs1}. The irreducible representations of
the
Poincar\'{e} group for massless particles
 \cite{wigner} can be constructed from a knowledge of the little group
 $G$ of a light-like
momentum four-vector $p = (p^{0}, {\bf p})$ . This group is the Euclidean group
$E(2)$. Physically, we are interested in possible finite-dimensional
representations of the covering of the little group.
We therefore restrict ourselves to the compact subgroup, i.e. we represent the
$E(2)$-translations trivially and
consider $G=SO(2)=U(1)$. Since the origin in the momentum space is
excluded for massless particles  one is therefore led to consider
$G$-bundles over $S^2$ since the energy of the particle can be kept
fixed. Such $G$-bundles are classified by mappings from the equator to
G, i.e. by the first homotopy group $\Pi_{1} (U(1))$={\bf Z}, where it turns
out that each integer
corresponds to twice the helicity of the particle. A massless particle
with helicity $\lambda $ and {\it sharp momentum} is thus  described in
terms of a non-trivial line bundle characterized by $\Pi _{1}(U(1)) = \{
2\lambda \} $ \cite{nagel}.

This consideration can easily be extended to higher space-time dimensions.
If $D$ is the number of space-time
dimensions, the corresponding $G$-bundles are classified by the homotopy groups
$\Pi _{D-3}
(Spin(D-2))$. These homotopy groups are in general non-trivial. It is
remarkable fact that the only trivial homotopy groups  of this form in higher
space-time
dimensions correspond to $D=5$ and $D=9$  due to the existence of quaternions
and the Cayley  numbers (see e.g. Ref. \cite{white}).  In these space-time
dimensions, and for $D=3$, it then turns that one can explicitly construct
canonical {\it and}
commuting position operators for massless particles \cite{albs1}. The
mathematical fact that the spheres $S^{1}$, $S^{3}$ and $S^{7}$ are
parallelizable can then be expressed in terms of the existence of canonical
{\it and} commuting position operators for massless spinning particles in
$D=3$, $D=5$ and $D=9$ space-time dimensions.

 In terms of a
canonical momentum $p_{i}$ and coordinates $x_{j}$ satisfying the canonical
commutation relation Eq.(\ref{ccr}) we can easily derive
the commutator of two components of the position operator ${\bf x}$ by making
use of a
simple consistency argument as follows.  If the massless
particle has a given helicity $\lambda$, then the generators of angular
momentum is given by:
 \begin{equation}
J_{k} = \epsilon _{klm}x_{l}p_{m} +
\lambda \frac{p_{k}}{|{\bf p}|}~~~.\label{eqn2}
\end{equation}
Tha canonical momentum then transforms as a vector under rotations, i.e.
\begin{equation}
[J_{k},p_{l}] = i\epsilon _{klm}p_{m}~~~,
\end{equation}
without any condition on the commutator of two components of the position
operator {\bf x}. The
 position operator will, however, not transform like a vector
unless the following commutator is postulated
\begin{equation}
i[x_{k},x_{l}] = \lambda \epsilon _{klm}\frac{p_{m}}{|{\bf p}|^{3}}~~~,
\label {eqn4}
 \end{equation}
where we notice that commutator formally corresponds to a pointlike Dirac
magnetic
monopole \cite{dirac} localized at the origin in momentum space with strength
$4\pi
\lambda$. The energy $p^{0}$ of the massless particle is, of course, given by
 $\omega = |{\bf p}|$. In terms of a singular $U(1)$ connection ${\cal
A}_{l}\equiv {\cal A}_{l}({\bf p})$ we can write
\begin{equation}
x_k = i\partial _k - {\cal A }_{k}~~~,
\label{dcon}
\end{equation}
where $\partial _{k} =\partial /\partial p_{k} $ and
\begin{equation}
\partial _{k}{\cal A}_{l} - \partial _{l}{\cal A}_{k} = \lambda \epsilon
_{klm}\frac{p_{m}}{|{\bf p }|^{3}}~~~.
\end{equation}
Out of the observables $x_{k}$ and the energy $\omega$ one can
easily construct the generators (at time $t=0$) of Lorentz boots, i.e.
\begin{equation}
K_{m}= (x_{m}\omega + \omega x_{m})/2~~~~,
\label{boost}
\end{equation}
and verify that $J_{l}$ and $K_{m}$ lead to a
realization of the Lie algebra of the Lorentz group, i.e.
\begin{eqnarray}
&[J_{k},J_{l}]&=\, i\epsilon _{klm}J_{m}~~~~,\\
&[J_{k},K_{l}]&=\, i\epsilon _{klm}K_{m}~~~~,\\
&[K_{k},K_{l}]&=\, -i\epsilon_{klm}J_{m}~~~~.\end{eqnarray}
The components of the Pauli-Plebanski operator $W_\mu$ are given by
\begin{equation}
W^{\mu} = (W^{0},{\bf W}) = ({\bf J}\cdot {\bf p}, {\bf J}p^{0} +
{\bf K}\times {\bf p}) = \lambda p^{\mu}~~~,
\end{equation}
i.e. we also obtain an irreducible representation of the Poincar\'{e} group.
The additional non-zero commutators are
\begin{eqnarray}
& [K_{k},\omega ] &= ip_{k}~~~~, \\
& [K_{k},p_{l}] &=i\delta _{kl}\omega ~~~~.
\end{eqnarray}
At $t\equiv x^{0}(\tau )\neq 0$ the Lorentz boost generators $K_{m}$ as
given by Eq.(\ref{boost}) are extended to
\begin{equation}
K_{m}= (x_{m}\omega + \omega x_{m})/2 - tp_{m}~~~~.
\label{Kt}
\end{equation}
In the Heisenberg picture, the quantum equation of motion of an
observable ${{\cal O}(t)}$ is obtained by using
\begin{equation}
\frac{d{\cal O}(t)}{dt} = \frac{\partial {\cal O}(t)}{\partial t}
+ i[H,{\cal O}(t)]~~~~,
\end{equation}
where the Hamiltonian $H$ is given by the $\omega $. One then finds that
all generators of the Poincar\'{e} group are conserved as they should. The
equation of motion for {\bf x}(t) is
\begin{equation}
\frac{d}{dt} {\bf x}(t) = \frac{{\bf p}}{\omega}~~~~,
\end{equation}
which is an expected equation of motion for a massless particle.

The  non-commuting components $x_{k}$ of the position operator ${\bf x}$
transform
as the components of a vector under spatial rotations. Under Lorentz boost we
find in
addition that
\begin{equation}
i[K_{k},x_{l}] = \frac{1}{2}\left( x_{k}\frac{p_{l}}{\omega}
+ \frac{p_{l}}{\omega}x_{k} \right) - t\delta _{kl} +
\lambda \epsilon _{klm}\frac{p_m}{|{\bf p }|^2}~~~~.
\label{kx}
\end{equation}
The first two terms in Eq.(\ref{kx}) corresponds to the correct limit for
$\lambda = 0$ since the proper-time condition $x^{0}(\tau ) \approx \tau $ is
not Lorentz invariant \cite{hanson}. The last term in Eq.(\ref{kx}) is due to
the non-zero commutator Eq.(\ref{eqn4}). This anomalous term can be dealt with
by introducing an appropriate two-cocycle for finite transformations consisting
of translations generated by the position operator ${\bf x}$, rotations
generated by ${\bf J}$ and Lorentz boost generated by ${\bf K}$. For pure
translations this two-cocycle will be explicitly constructed in Section 3.

The
algebra discussed above can be extended in a rather straightforward manner
to incorporate both positive and negative helicities  needed in order to
describe e.g. linearly polarized light. As we now will see this extension
corresponds to a replacement of the Dirac monopole in momentum space with
a $SU(2)$ Wu-Yang \cite{wuyang} monopole. The procedure below follows a rather
standard method of imbedding the singular $U(1)$ connection ${\cal A}_{l}$ into
a {\it regular} $SU(2)$ connection.
 Let us specifically consider a massless, spin-one particle.
The Hilbert space, ${\cal H}$, of  one-particle transverse wave-functions
$\phi_{\alpha }({\bf p}), \alpha = 1,2,3$ is defined in terms of a scalar
product
\begin{equation}
(\phi,\psi) = \int d^{3}p \phi ^{*}_{\alpha }({\bf p})\psi _{\alpha}({\bf
p})~~~,
\end{equation}
where $\phi ^{*}_{\alpha }({\bf p})$ denotes the complex conjugated
 $\phi _{\alpha}({\bf p})$.
In terms of a Wu-Yang connection ${\cal A}^{a}_{k }\equiv {\cal A}^{a}_{k
}({\bf p})$, i.e.
\begin{equation}
{\cal A}^{a}_{k }({\bf p})= \epsilon _{alk}\frac{p_{l}}{|{\bf
p}|^{2}}~~~, \end{equation}
Eq.(\ref{dcon}) is extended to
\begin{equation}
x_k = i\partial _k - {\cal A }_{k}^{a}({\bf p})S_{a}~~~,
\label{wupos}
\end{equation}
where
\begin{equation}
(S_{a})_{kl}= -i\epsilon _{akl}
\end{equation}
 are the spin-one generators. By means of a singular gauge-transformation the
Wu-Yang connection can be transformed into the singular $U(1)$-connection
${\cal A}_{l}$ times the third component of the spin generators $S_{3}$.
This position operator defined by Eq.(\ref{wupos}) is compatible with the
transversality condition
on the one-particle wave-functions, i.e. $x_{k}\phi _{\alpha}({\bf p})$ is
transverse. With suitable conditions on the one-particle wave-functions the
position operator ${\bf x}$ therefore has a welldefined action on ${\cal H }$.
Furthermore,
\begin{equation}
i[x_{k},x_{l}] = {\cal
F}^{a}_{kl}S^{a} =  \epsilon _{klm}\frac{p_{m}}{|{\bf p}|^{3}}
\hat{{\bf p}}\cdot {\bf S}~~~,
\label{eq:gen}
 \end{equation}
where
\begin{equation}
{\cal F}^{a}_{kl} = \partial _{k}{\cal A}^{a}_{l}
 - \partial _{l}{\cal A}^{a}_{k}
-\epsilon _{abc}{\cal A}^{b}_{k}{\cal
A}^{c}_{l} = \epsilon _{klm}\frac{p_{m}p_{a}}{|{\bf p }|^{4}}~~~,
\end{equation}
is the non-Abelian $SU(2)$ field strength tensor and $\hat{{\bf p}}$ is a unit
vector in the direction of the particle
momentum ${\bf p}$. The generators of angular momentum are now defined as
follows
 \begin{equation}
J_{k} = \epsilon _{klm}x_{l}p_{m} +
\frac{p_{k}}{|{\bf p}|}\hat{{\bf p}}\cdot {\bf S}~~~.
\label{eq:ang}
\end{equation}
The helicity operator $ \Sigma \equiv \hat{{\bf p}}\cdot {\bf S}$ is
covariantly constant, i.e.
\begin{equation}
\partial _{k} \Sigma + i\left[A_{k},\Sigma \right] = 0~~~,
\label{ccon}
\end{equation}
where $A_{k} \equiv {\cal A}_{k}^{a}({\bf p})S_{a}$. The position operator
${\bf x}$ therefore commutes with  $\hat{{\bf p}}\cdot {\bf S}$. One can
therefore verify in a straightforward manner that the observables $p_{k},
\omega  , J_{l}$ and $K_{m} = (x_{m}\omega +\omega x_{m})/2$ close to the
Poincar\'{e} group. At $t \neq 0$ the Lorentz boost generators $K_{m}$ are
defined as in Eq.(\ref{Kt}) and Eq.(\ref{kx}) is extended to
\begin{equation}
i[K_{k},x_{l}] = \frac{1}{2}\left( x_{k}\frac{p_{l}}{\omega}
+ \frac{p_{l}}{\omega}x_{k} \right) - t\delta _{kl} +
i\omega[x_{k},x_{l}]~~~~.
\label{kx2}
\end{equation}

For helicities $\hat{{\bf p}}\cdot {\bf S}= \pm \lambda$ one
extends the previous considerations
 by considering ${\bf S}$ in the spin $|\lambda |$-representation.
Eqs.(\ref{eq:gen}), (\ref{eq:ang}) and (\ref{kx2}) are then valid in general. A
reducible representation for the generators
of the Poincar\'{e} group for an arbitrary spin has therefore been constructed
for a massless particle. We observe that the helicity operator $\Sigma$ can be
interpreted as a
generalized ``magnetic charge'', and since $\Sigma$ is covariantly conserved
one can use the general theory of topological quantum numbers \cite{goddard}
and derive the quantization condition
\begin{equation}
\exp (i4\pi \Sigma ) =1~~~~,
\end{equation}
i.e. the helicity is properly quantized.

\begin{center}
\section{\sc Topological Spin}
\end{center}
\setcounter{equation}{0}
Coadjoint orbits on a group $G$  admit a symplectic two-form (see e.g.
\cite{kir}) which can be used to construct topological Lagrangians, i.e.
Lagrangians constructed by means of Wess-Zumino terms \cite{wesszumino} (for a
general account see e.g. \cite{classical}).
Let us
illustrate the basic ideas for a non-relativistic spin and $G = SU(2)$.
Let ${\cal K}$ be an element of the Lie algebra ${\cal G}$ of $G$ in the
fundamental representation.
Without loss of generality we can write ${\cal K} = \lambda _{\alpha}\sigma
_{\alpha} = \lambda \sigma _3$, where $\sigma _{\alpha}, \alpha = 1,2,3$
denotes the three Pauli spin matrices. Let $H$ be the little
group of ${\cal K}$. Then the coset space $G/H$ is isomorphic to $S^2$ and
defines an adjoint
orbit (for semi-simple Lie groups adjoint and coadjoint representations are
equivalent due to the existence of the non-degenerate Cartan-Killing form).
The action for the spin degrees of freedom is then expressed in terms of the
group $G$ itself, i.e.
\begin{equation}
S_{P} = -i\int \bk{{\cal K}}{g^{-1}(\tau )dg(\tau )/d\tau} d\tau ~~~~,
\end{equation}
where $\bk{A}{B}$ denotes the trace-operation of two
Lie-algebra elements $A$ and $B$ in ${\cal G}$ and where
\begin{equation}
g(\tau )= \exp (i\sigma
_{\alpha} \xi _{\alpha}(\tau ))
\end{equation}
 defines the (proper-)time dependent
dynamical group element.
 We observe that $S_{P}$ has
a gauge-invariance, i.e. the transformation
\begin{equation}
g(\tau ) \longrightarrow g(\tau ) \exp \left( i\theta (\tau )\sigma _3
\right) \label{uone}
\end{equation}
only change the Lagrangian density $\bk{{\cal K }}{g^{-1}(\tau )dg(\tau )/d\tau
}$ by a total time
derivative. The gauge-invariant components of spin, $S_{k}(\tau )$,
 are defined in terms of ${\cal K}$
by the relation
 \begin{equation}
 S(\tau ) \equiv S_{k}(\tau )\sigma _{k} = \lambda
g(\tau )\sigma _{3} g^{-1}(\tau ) ~~~,
\label{spindef}
 \end{equation}
such that
\begin{equation}
S^2 \equiv S_{k}(\tau )S_{k}(\tau ) = \lambda ^{2}~~~.
\end{equation}
By adding a non-relativistic particle kinetic term as well as  a conventional
magnetic moment
interaction term to the action $S_{P}$, one can verify that the components
$S_{k}(\tau )$
 obey the correct
classical equations of motion for spinprecession \cite{baletal,classical}.

Let $M = \{ \sigma ,\tau | \sigma \in [0,1] \}$ and
 $ (\sigma ,\tau ) \rightarrow g(\sigma ,\tau )$ parametrize $\tau
-$dependent paths in $G$ such that $g(0 ,\tau ) = g_{0}$ is an arbitrary
reference element and $g(1,\tau ) = g(\tau )$. The Wess-Zumino term in this
case is given by
\begin{equation}
 \omega _{WZ} = -id \bk{{\cal
K}}{g^{-1}(\sigma, \tau )dg(\sigma, \tau )} =
 i\bk{{\cal K}}{(g^{-1}(\sigma, \tau )dg(\sigma, \tau ))^{2}}~~~,
  \end{equation} where
$d$ denotes exterior differentiation and where now
\begin{equation}
g(\sigma, \tau )= \exp (i\sigma
_{\alpha} \xi _{\alpha}(\sigma, \tau ))~~~~.
\end{equation}
 Apart from boundary terms which do not contribute to the equations of
motion, we then have that \begin{equation}
S_{P} = S_{WZ} \equiv  \int _{M}\omega _{WZ} =
-i\int _{\partial M} \bk{{\cal K}}{g^{-1}(\tau )dg(\tau )}~~~,
\label{wzaction}
\end{equation}
where the one-dimensional boundary $\partial M $ of $M$ , parametrized
by $\tau $, can play the role of (proper-) time. $\omega _{WZ}$ is now
gauge-invariant under a larger $U(1)$ symmetry, i.e. Eq.(\ref{uone}) is now
extended to
\begin{equation}
g(\sigma,\tau ) \longrightarrow g(\sigma,\tau ) \exp \left( i\theta
(\sigma,\tau )\sigma _3 \right)~~~~.
\end{equation}
$\omega _{WZ}$ is therefore a two-form defined on the coset space $G/H$.
A canonical analysis then shows that there are no gauge-invariant dynamical
degrees of freedom in the interior of $M$.
The Wess-Zumino action Eq.(\ref{wzaction}) is the
topological action for spin degrees of freedom.

As for the quantization of the theory described by the action
Eq.(\ref{wzaction}), one may use methods from geometrical quantization and
especially the Borel-Weil-Bott theory of representations of compact Lie
groups \cite{kir,classical}. One then finds that $\lambda$ is half an
integer, i.e. $|\lambda |$ corresponds to the spin. This quantization of
$\lambda $ also
naturally emerges by demanding that the action Eq.(\ref{wzaction}) is
welldefined in
quantum mechanics for  periodic motion as recently was discussed by e.g.
Klauder \cite{klauder}, i.e.
\begin{equation}
4\pi \lambda = \int _{S^{2}} \omega _{WZ} = 2\pi n ~~~~,
\end{equation}
where $n$ is an integer. The symplectic two-form $\omega _{WZ}$ must then
belong to an integer class cohomology.
 This geometrical approach is in principal
straightforward, but it requires explicit coordinates on $G/H$. An
alternative approach, as used in \cite{baletal,classical}, is a
canonical Dirac analysis and quantization \cite{hanson}. This procedure leads
to the
condition $\lambda ^{2} = s(s+1)$, where $s$ is half an integer. The
fact that one can arrive at different answers for $\lambda $ illustrates
a certain lack of uniqueness in the quantization procedure of the action
Eq.(\ref{wzaction}). The quantum theories obtained describes, however, the
same physical system namely one irreducible representation of the group
$G$.

The action Eq.(\ref{wzaction}) was first proposed in \cite{bor}. The action can
be derived quite naturally in terms of a coherent state path integral (for a
review see e.g. Ref.\cite{klauderme}) using spin coherent states. It is
interesting to notice that structure of the action Eq.(\ref{wzaction}) actually
appears in such a language already in a
paper by Klauder on continuous representation theory \cite{klauderaction}.

A classical action which after quantization leads to a description of a
massless particle in terms of an irreducible representations of the
Poincar\'{e}
group can be constructed in a similar fashion \cite{baletal}. Since the
Poincar\'{e} group is non-compact the geometrical analysis referred to above
for
non-relativistic spin must be extended and one should consider coadjoint
orbits instead of adjoint orbits (D=3 appears to be an
exceptional case due to the existence of a non-degenerate bilinear form
on the D=3 Poincar\'{e} group Lie algebra \cite{witten}. In this case there
is a topological action for irreducible representations of the form
Eq.(\ref{wzaction}) \cite{topq}). The action then takes the form
\begin{equation}
 S= \int d\tau \left( p_{\mu}(\tau )\dot{x}^{\mu}(\tau
) + \frac{i}{2} \mbox{Tr} [{\cal K } \Lambda ^{-1}(\tau )
\frac{d}{d\tau}\Lambda (\tau )]
 \right)~~~.
\label{paction}
\end{equation}
Here $[\sigma _{\alpha \beta}]_{\mu \nu} = -i(\eta _{\alpha \mu}\eta _{\beta
\nu}
-\eta _{\alpha \nu } \eta_{\beta \mu})$ are the Lorentz group generators in the
spin one representation and $\eta _{\mu \nu} = (-1,1,1,1)$ is the Minkowski
metric. The Lorentz group Lie-algebra element ${\cal K }$ is chosen to be
$\lambda \sigma _{12}$. The $\tau $-dependence of the Lorentz group element
$\Lambda _{\mu \nu}(\tau )$ is defined by
\begin{equation}
\Lambda _{\mu \nu}(\tau ) = \left[ \exp \left(i\sigma _{\alpha \beta} \xi
^{\alpha \beta }(\tau ) \right)\right]_{\mu \nu}~~~~.
\end{equation}
 The momentum variable $p_{\mu}(\tau )$ is defined by
\begin{equation}
p_{\mu }(\tau ) = \Lambda _{\mu \nu} (\tau ) k^{\nu }~~~,
\end{equation}
where the constant reference momentum $ k^{\nu }$ is given by
\begin{equation}
k^{\nu }
 = (\omega ,0,0,|{\bf k }|)~~~,
\end{equation}
where $\omega = |{\bf k }|$. The momentum $p_{\mu }(\tau )$ is the light-like
by construction. The action Eq.(\ref{paction}) leads to the equations of motion
\begin{equation}
\frac{d}{d\tau}p_{\mu}(\tau ) = 0~~~,
\end{equation}
and
\begin{equation}
\frac{d}{d\tau } \left\{ x_{\mu }(\tau )p_{\nu}(\tau ) - x_{\nu}(\tau
)p_{\mu}(\tau ) + S_{\mu \nu }(\tau ) \right\} = 0 ~~~.
\end{equation}
Here we have defined gauge-invariant spin degrees of freedom $ S_{\mu \nu
}(\tau ) $ by
\begin{equation}
 S_{\mu \nu }(\tau )  = \frac{1}{2} \mbox{Tr}[\Lambda (\tau ){\cal K }\Lambda
^{-1}(\tau )\sigma _{\mu\nu}]
\end{equation}
in analogy with Eq.(\ref{spindef}). These spin degrees of freedom satisfy
the relations
\begin{equation}
p_{\mu} (\tau ) S^{\mu \nu }(\tau ) = 0~~~,
\label{CON1}
\end{equation}
and
\begin{equation}
\frac{1}{2} S_{\mu \nu }(\tau ) S^{\mu \nu }(\tau ) = \lambda ^{2}~~~.
\label{CON2}
\end{equation}
Inclusion of external electromagnetic and gravitational fields
leads to the classical Bargman-Michel-Telegdi and Papapetrou equations of
motion respectively \cite{baletal}. Since the equations derived are
expressed in terms of {\it bosonic} variables these equations of motion
admit a straightforward classical interpretation. (An alternative {\it bosonic}
treatment of internal degrees of freedom can be found in Ref.\cite{bssw}.)

Canonical quantization of the system described by {\it bosonic} degrees of
freedom and the action Eq.(\ref{paction}) leads to a realization of the
Poincar\'{e} Lie algebra with generators $p_{\mu}$ and $J_{\mu \nu}$ where
\begin{equation}
J_{\mu \nu} = x_{\mu}p_{\nu} -  x_{\nu}p_{\mu} + S_{\mu \nu}~~~~.
\end{equation}
The four vectors $x_{\mu}$ and $p_{\nu}$ commute with the spin generators
$S_{\mu \nu}$ and are canonical, i.e.
\begin{eqnarray}
&[x_{\mu},x_{\nu}] &=\,  [p_{\mu},p_{\nu}] = 0~~~~, \\
&[x_{\mu},p_{\nu}] &=\, i\eta _{\mu \nu}~~~~.
\end{eqnarray}
The spin generators $S_{\mu \nu}$ fulfil the conventional algebra
\begin{equation}
[S_{\mu \nu},S_{\lambda \rho}] = i(\eta _{\mu \lambda}S_{\nu \rho} +
\eta _{\nu \rho}S_{\mu \lambda} - \eta _{\mu \rho}S_{\nu \lambda}
-\eta _{\nu \lambda}S_{\mu \rho})~~~~.
\end{equation}
The mass-shell condition $p^{2}=0$ as well as the constraints Eq.(\ref{CON1})
and Eq.(\ref{CON2}) are all first-class constraints \cite{hanson}. In the
proper-time gauge $x^{0}(\tau ) \approx \tau $ one obtains the system described
in Section 2, i.e. we obtain an irreducible representation of the Poincar\'{e}
group with helicity $\lambda $ \cite{baletal}. For half-integer helicity, i.e.
for fermions, one can verify in a straightforward manner that the
wave-functions obtained change with a minus-sign under a $2\pi $
rotation \cite{baletal,albs1,classical} as they should.
\begin{center}
\section{\sc The Berry Phase for Photons}
\end{center}
\setcounter{equation}{0}

We have constructed a set of $O(3)$-covariant position  operators of massless
particles corresponding to a reducible representation of the Poincar\'{e} group
corresponding to a combination of positive and negative helicity. It is
interesting to notice that the
construction above leads to observable effects. Let us specifically consider
photons and the motion of photons along an optical fibre. Berry has argued
\cite{berry} that a spin in an adiabatically changing magnetic field leads
to the appearance of an observable phase factor, called the Berry phase. It
was suggested \cite{wu} that a similar geometric phase could appear for
photons. We will now,  within the framework of {\it relativistic quantum
mechanics}, provide for a derivation of this geometrical phase in terms of the
operator realization of the Poincar\'{e} discussed above. The Berry phase for
photons can then be obtained as follows. We consider the motion of a
photon with fixed energy moving in an optical fibre. We assume that as  the
photon moves in the fibre, the momentum vector traces out a closed loop in
momentum space on the constant energy surface, i.e. on a two-sphere $S^2$. We
therefore consider wave-functions $\ket{\bf p}$ which are diagonal in
momentum. We also define the translation operators  $U({\bf a}) = \exp (i{\bf
a}\cdot {\bf x})$. It is straightforward to show using  Eq.(\ref{eqn4}) that
 \begin{equation} U({\bf a})U({\bf b}) \ket{\bf b} = \exp
(i\gamma [{\bf a},{\bf b};{\bf p}]) \ket{{\bf p}+{\bf a}+{\bf b}}~~~~,
\end{equation}
where the two-cocycle phase $\gamma [{\bf a},{\bf b};{\bf p}]$ is equal
to the flux of the magnetic monopole in momentum space through the simplex
spanned by the vectors ${\bf a}$ and ${\bf b}$ localized at  the point ${\bf
p}$, i.e.
 \begin{equation} \gamma [{\bf a},{\bf b};{\bf p}]
 = \lambda \int_{0}^{1} \int_{0}^{1} d\xi _{1} d\xi _{2} a_{k}b_{l}
\epsilon _{lkm} B_{m}({\bf p} + \xi _{1} {\bf a} + \xi _{2} {\bf b})~~~~,
\end{equation}
where $B_{m}({\bf p}) = p_{m}/|{\bf p}|^{3}$. The non-trivial phase appears
because the second de Rham cohomology group of $S^{2}$ is non-trivial. The
two-cocycle phase $\gamma [{\bf a},{\bf b};{\bf p}]$ is therefore not a
coboundary and hence it cannot be removed by a redefintion of $U({\bf a})$.
This result
has a close analogy in the theory of
magnetic monopoles \cite{jackiw}. The anomalous commutator Eq.(\ref{eqn4})
therefore leads to a ray-representation of the translations in momentum space.

A closed loop
in momentum space, starting
and ending at ${\bf p}$, can then be obtained by using a sequence of
infinitesimal translations
 $U(\delta {\bf a})\ket{{\bf p}} = \ket{{\bf p}+ \delta{\bf a}}$ such that
$\delta {\bf a}$ is orthogonal to argument of the wave-function on which it
acts (this defines the adiabatic transport of the system). The momentum
vector ${\bf p}$ then traces out a closed curve on the constant energy
surface $S^2$ in momentum space.  The total phase of these translations then
gives a phase $\gamma $ which is the $\lambda $ times the solid angle of the
closed curve the momentum vector traces out on the constant energy surface.
This phase does not depend on Plancks constant.
This is precisely the Berry phase for the photon with a given helicity
$\lambda$. In the experiment by Tomita and Chiao \cite{chiao} one considers a
linearly polarized photon. The same line of arguments above but making use
Eq.(\ref{eq:gen}) instead of  Eq.(\ref{eqn4}) leads to the desired change of
polarization as the photon moves along the optical fibre. An alternative
derivation of the Berry phase for photons is based on observation that the
covariantly conserved helicity operator $\Sigma$ can be interpreted as a
generalized ``magnetic charge''. Let $\Gamma$ denote a closed path in momentum
space parametrized by $\sigma \in [0,1]$ such that ${\bf p}(\sigma =0) = {\bf
p}(\sigma = 1) = {\bf p}_{0}$ is fixed. The parallel transport of a
one-particle state $\phi _{\alpha}({\bf p})$ along the path $\Gamma$ is then
determined by a path-ordered exponential, i.e.
\begin{equation}
\phi _{\alpha}({\bf p}_{0}) \longrightarrow  \left[ P\exp \left(i\int
_{\Gamma}A_{k}({\bf p}(\sigma ))\frac{dp_{k}(\sigma )}{d\sigma}d\sigma \right)
\right]_{\alpha \beta}\phi _{\beta} ({\bf p}_{0})~~~~,
\end{equation}
where $A_{k}({\bf p}(\sigma))  \equiv {\cal A}_{k}^{a}({\bf p}(\sigma ))S_{a}$.
By making use of a non-Abelian version of Stokes theorem \cite{goddard} one can
then show that
\begin{equation}
 P\exp \left(i\int _{\Gamma}A_{k}({\bf p}(\sigma ))\frac{dp_{k}(\sigma
)}{d\sigma}d\sigma \right) = \exp\left(i\Sigma \Omega [\Gamma ]\right)~~~~,
\end{equation}
where $\Omega [\Gamma ]$ is the solid angle subtended by the path $\Gamma$ on
the two-sphere $S^{2}$. This result leads again to the desired change of linear
polarization as the photon moves along the path described by $\Gamma$. This
derivation does not require that $|{\bf p}(\sigma )|$ is constant along the
path.

In the experiments so far considered the photon flux is large. In order to
strictly apply our results under such conditions one can consider a second
quantized version of the theory we have presented following e.g. the discussion
of Amrein \cite{jpa}. By making use of coherent states of the electromagnetic
field in a standard and straightforward manner (see e.g. Ref.\cite{klauderme})
one then realize that our considerations survive. This is so since the coherent
states are parametrized in terms of the one-particle states. By construction
the coherent states then inherits the transformation properties of the
one-particle states discussed above.
\begin{center}
\section{\sc Final Remarks and Conclusions}
\end{center}
\setcounter{equation}{0}

In the analysis of Wightman, corresponding to commuting position
variables, the natural mathematical tool turned out to be systems of
imprimitivity for the representations of the three-dimensional Euclidean
group. In the case of non-commuting position operators we have also seen
that notions from differential geometry are important. It is interesting
to see that such a broad range of mathematical methods enters into the
study of the notion of localizability of physical systems.
 We have in particular argued that
Abelian as well as non-Abelian magnetic monopole field configurations reveal
themselves in a description of localizability of
massless spinning particles.
  Concerning the physical existence of magnetic monopoles
 Dirac remarked in 1981 \cite{craige} that ``{\it I am inclined now
to believe that monopoles do not exist. So many years have gone by without
any encouragement from the experimental side}''. The
``monopoles'' we are considering appear, however, as mathematical objects
in the momentum space of the massless particles. Their existence, we have
argued, is then only indirectly revealed to us by the properties of e.g.
the photons moving along optical fibres.

Localized states of massless particles will necessarily develop
non-exponential tails in space as a consequence of the Hegerfeldts theorem
\cite{heger2}. Various  number operators representing the number of
massless, spinning particles localized in a finite volume $V$ at time $t$
has been discussed in the literature. The non-commuting position
observables we have discussed for photons correspond to the pointlike
limit of the weak localizability of Jauch, Piron and Amrein \cite{jpa}.
This is so since our construction, as we have seen in Section 2,
corresponds to an explicit enforcement of the transversality condition of
the one-particle wave-functions.

In a finite volume, photon number operators appropriate for weak
localization \cite{jpa} do not agree with the photon number operator
introduced by Mandel \cite{mandel1} for sufficiently small wavelengths as
compared to the linear dimension of the localization volume. It would be
interesting to see if there are measurable differences. A necessary
ingredient in answering such a question would be the experimental
realization of a localized one-photon state. It is interesting to notice
that such states can be generated in the laboratory \cite{mandel2}.

In concluding we find it appealing  that the quantization of
the system describing a photon, in general linearly polarized, is ``{\sl
geometry, after all~}'' \cite{klauder}.
\vspace{3mm}
 \begin{center} {\bf
ACKNOWLEDGEMENT} \end{center} \vspace{3mm}
We are greatful for a fruitful collaboration with
A. P. Balachandran, G. Marmo and A. Stern on projects related to the subject in
question. We also acknowledge discussions with M. Berry and D. Kastler. I am
also very
greatful to John R. Klauder for sharing with him his deep insights in
theoretical
physics in a spirit of friendship and search for truth.


\begin{thebibliography}{999}
%
\bibitem{newton} T. Newton and E.P. Wigner,  ``{\sl Localized States for
Elementary Systems}'', Rev. Mod Phys. {\bf 21} (1949) 400.
%
\bibitem{wightman}
A. S. Wightman, ``{\sl On the Localizability
 of Quantum Mechanical Systems}'', Rev. Mod. Phys. {\bf 34}(1962) 845.
%
\bibitem{jpa}
J. M. Jauch and C. Piron, ``{\sl Generalized Localizability}'',
 Helv. Phys. Acta {\bf 40} (1967) 559;\\
W. O. Amrein, ``{\sl Localizability for Particles of Mass Zero}'', Helv.
Phys. Acta {\bf 42} (1969) 149.
%
\bibitem{local}
T. F. Jordan and N. Mukunda, ``{\sl Lorentz-Covariant Position Operators for
Spinning Particles}'', Phys. Rev. {\bf 132} (1963) 1842;\\
A. L. Licht, ``{\sl Local States}'',
 J. Math. Phys. {\bf 7} (1966) 1656;\\
K. Krauss, ``{\sl Position Observables of
the Photon}'' in ``{\sl The Uncertainty Principle and Foundations of Quantum
Mechanics}", Eds. W.C. Price and S.S. Chissick (John Wiley \& Sons, 1977);\\
M. I. Shirokov, ``{\sl Strictly Localized States and Particles}'', Theor.
Math. Phys. {\bf 42} (1980) 134;\\ S. N. M. Ruijsenaars, ``{\sl On
Newton-Wigner Localization and Superluminal Propagation Speeds}'', Ann. Phys.
(N.Y.) {\bf 137} (1981) 33;\\ E. Prugovecki, ``{\sl Stochastic Quantum
Mechanics and Quantum Spacetime}" (Kluwer Academic, Hingham, Mass., 1984).
 %
\bibitem{ehlers} J. Ehlers, A. E. Pirani and A. Schild, ``{\sl The Geometry of
Free-Fall and Light Propagation}'' in ``{\sl General Relativity. Papers in
Honor of J. L. Synge}'', Ed. L. O'Raifeartaigh (Oxford University Press,
London, 1972).
%
\bibitem{wigner} E. Wigner, ``{\sl Unitary Representations
of the Inhomogeneous Lorentz Group}'', Ann. Math. {\bf 40} (1939) 149.
 %
\bibitem{heger} G. C. Hegerfeldt, ``{\sl Remark on the Causality  and Particle
Localization}'', Phys. Rev. {\bf D10} (1974) 3320; \\  B.-S. Skagerstam,
``{\sl Some Remarks Concerning the Question of Localization of Elementary
Particles}'', Int. J. Theor. Phys. {\bf 15} (1976) 213;\\ J. F. Perez and I. F.
Wilde, {\sl Localiztion and Causality in Relativistic Quantum Mechanics}'',
Phys, Rev. {\bf D16} (1977) 315;\\ G. C. Hegerfeldt and S. N. M. Ruijsenaars,
``{\sl Remarks on Causality, Localization and Spreading of Wave Packets}'',
Phys. Rev. {\bf D22} (1980) 377.
 %
\bibitem{heger2} G. C. Hegerfeldt, ``{\sl Violation of Causality in
Relativistic Quantum Theory?}'', Phys. Rev.
Lett. {\bf 54} (1985) 2395.
%
\bibitem{baletal}  A. P. Balachandran, G. Marmo, B.-S.  Skagerstam and A.
Stern, ``{\sl Spinning Particles in General Relativity}'', Phys. Lett. {\bf
89B}
(1980) 199 and ``{\sl Gauge Symmetries and Fibre Bundles: Applications to
Particle Dynamics}" , Lecture Notes in Physics Vol. {\bf 188}
(Springer Verlag, 1983); \\ B.-S. Skagerstam and A. Stern, ``{\sl Lagrangian
Description of Classical Charged Particles with Spin}'', Physica Scripta {\bf
24} (1981) 493.
%
\bibitem{atre} M. Atre, A. P. Balachandran and T. V. Govindarajan, ``{\sl
Massless Spinning Particles in All Dimensions and Novel Magnetic
Monopoles}'', Int. J. Mod. Phys. {\bf A2} (1987) 453.
%
\bibitem{albs1} B.-S. Skagerstam and A. Stern, ``{\sl Light-Cone  Gauge
Versus Proper-Time Gauge for Massless Spinning Particles}'', Nucl. Phys. {\bf
B294} (1987) 636.
 %
\bibitem{nagel} B. Ek and B. Nagel, ``{\sl Differentiable Vectors and  Sharp
Momentum States of Helicity Representations of the Poincar\'{e} Group}'', J.
Math. Phys. {\bf 25} (1984) 1662.
 %
\bibitem{white} G. W. Whitehead, ``{\sl Elements of Homotopy Theory}''
(Springer Verlag, 1978).
%
\bibitem{dirac} P. A. M. Dirac ``{\sl Quantized Singularities in the
Electromagnetic Field}'', Proc. Roy. Soc. {\bf A133} (1931) 60; ``{\sl The
Theory of Magnetic Monopoles}'', Phys. Rev. {\bf 74} (1948) 817.
 %
\bibitem{hanson} A. J. Hansson, T. Regge and C. Teitelboim, ``{\sl
Constrained Hamiltonian Systems}'' (Accademia Nazionale dei Lincei, Roma
1976).
 %
 \bibitem{wuyang} T. T. Wu and C. N. Yang, ``{\sl Some Remarks
About Unquantized Non-Abelian Gauge Fields}'', Phys. Rev. {\bf D12} (1975)
3843.
 %
\bibitem{goddard} P. Goddard, J. Nuyts and D. Olive, ``{\sl Gauge Theories and
Magnetic Charge}'', Nucl. Phys. {\bf B125} (1977) 1.
%
 \bibitem{kir}  A. A.
Kirillov,  ``{\sl    Elements of the Theory of Representations~}'', A Series of
Comprehensive Studies in Mathematics {\bf 220} (Springer Verlag, Berlin 1976).
For some
recent applications see e.g. E. Witten, ``{\sl Coadjoint Orbits of the Virasoro
Group}'', Commun. Math. Phys. {\bf 114} (1988) 1;\\ A. Aleksev, L.
Fadeev and S. Shatashvili, ``{\sl Quantization of the Symplectic Orbits of the
Compact
Lie Groups by Means of the Functional Integral~}'', J. Geom. Phys. {\bf 1}
(1989) 3;\\
A. Aleksev and S. Shatashvili,  ``{\sl    Path Integral Quantization of the
Coadjoint
Orbits of the Virasoro Group and 2D Gravity~}'', Nucl. Phys. {\bf B323} (1989)
719;\\ O. Alvarez, I. M. Singer and P.
Windey, ``{\sl Quantum Mechanics and the Geometry of the Weyl Character
Formula~}'',
Nucl. Phys. {\bf B337} (1990) 467;\\
B. Rai and V. G. J. Rodgers,  ``{\sl    From Coadjoint Orbits to Scale
Invariant
WZNW Type Actions and 2-D Quantum Gravity Action~}'', Nucl. Phys.
{\bf B341} (1990) 119;\\
G. W. Delius, P. van Nieuwenhuizen and V. G. J. Rodgers,
``{\sl The Method of Coadjoint
Orbits: An Algorithm for the Construction of Invariant Actions~}'',
Int. J. Mod. Phys. {\bf A5} (1990) 3943.
%
\bibitem{wesszumino} J. Wess and B. Zumino, ``{\sl Consequences of Anomalous
Ward Indentities}'', Phys. Lett. {\bf 37B} (1971) 95.
 %
\bibitem{classical}  A. P. Balachandran, G. Marmo, B.-S.  Skagerstam and A.
Stern, ``{\sl Classical Topology and Quantum States}"  (World Scientific,
1991).
%
\bibitem{klauder}
J. R. Klauder, ``{\it Quantization {\it Is} Geometry, after All~}'', Ann.
Phys. (N.Y.) {\bf 188} (1988) 120.
%
\bibitem{bor} A. P. Balachandran, S. Borchardt and A. Stern, ``{\sl
Lagrangian and Hamiltonian Descriptions of Yang-Mills Particles}'', Phys.
Rev. {\bf D11}(1978) 3247.
%
\bibitem{klauderme} J. R. Klauder and B.-S. Skagerstam, ``{\sl Coherent
States - Applications in Physics and Mathematical Physics}'' (World
Scientific, Singapore 1985 and Beijing 1988).
%
 \bibitem{klauderaction} J. R. Klauder, ``{\sl
Continuous-Representation Theory.II. Generalized Relation Between Quantum and
Classical Dynamics.}'', J. Math. Phys. {\bf 4}(1963) 1058.
%
\bibitem{witten} E. Witten, ``{\sl 2 + 1 Dimensional Gravity As An Exactly
Solvable System}'', Nucl. Phys. {\bf B311 }(1988) 46.
%
\bibitem{topq} B.-S. Skagerstam and A. Stern, ``{\sl Topological Quantum
Mechanics in 2+1 Dimensions}'', Int. J. Mod. Phys. {\bf A5} (1990) 1575.
%
\bibitem{bssw} A. P. Balachandran, P. Salomonson, B.-S. Skagerstam and J.-O.
Winnberg, ``{\sl Classical Description of a Particle Interacting With a
Non-Abelian Gauge Field}'', Phys. Rev. {\bf D15} (1977) 2308.
%
 \bibitem{berry} M. V. Berry, ``{\sl
Quantal Phase Factors Accompanying Adiabatic Changes}'', Proc. Roy. Soc.
(London) {\bf 392} (1984) 45;\\ B. Simon, ``{\sl Holonomy, the Quantum
Adiabatic Theorem,  and Berry's Phase}'', Phys. Rev. Lett. {\bf 51} (1983)
2167. \\ For reviews see e.g. I. J. R. Aitchison, ``{\sl Berry Phases,
Magnetic Monopoles, and Wess-Zumino Terms or How the Skyrmion Got Its
Spin}'', Acta. Phys. Polon, {\bf B18} (1987) 207; ``{\sl Berry's Topological
Phase in Quantum Mechanics and Quantum Field Theory}'', Physica Scripta {\bf
T23} (1988) 12;\\ M. V. Berry, ``{\sl Quantum Adiabatic Anholonomy}" in
``{\sl Anomalies, Phases, Defects...}", Eds. M. Bregola, G. Marmo and G.
Morandi (Bibliopolis, Napoli, 1990);\\ S. I. Vinitski\u{i}, V. L. Derbov, V.
N. Dubovik,  B. L. Markovski and Yu P. Stepanovski\u{i}, ``{\sl Topological
Phases in Quantum Mechanics and Polariztion Optics}'', Sov. Phys. Usp. {\bf
33} (1990) 403.
 %
\bibitem{wu} R.Y. Chiao and Y.-S. Wu, ``{\sl Manifestations of Berry's
Topological Phase for the Photon}'', Phys. Rev. Lett. {\bf 57} (1986) 933.
%
\bibitem{jackiw} R. Jackiw, ``{\sl Three-Cocycle in Mathematics and
Physics}'', Phys. Rev. Lett. {\bf 54} (1985) 159; ``{\sl Magnetic Sources and
3-Cocycles (Comment)}'',  Phys. Lett. {\bf 154B} (1985) 303;\\ B. Grossman,
``{\sl A 3-Cocycle in Quantum Mechanics}'', Phys. Lett. {\bf 54} (1985) 159;
``{\sl Three-Cocycle in Quantum Mechanics. II}'', Phys. Rev. {\bf D33} (1986)
2922;\\ Y. S. Wu and A. Zee, ``{\sl Cocycles and Magnetic Monopole}'', Phys.
Lett. {\bf 152B} (1985) 98;\\
 J. Mickelsson,``{\sl Comment on
``Three-Cocycle in Mathematics and Physics''}'', Phys. Rev. Lett. {\bf
54} (1985) 2379;\\ D. Boulware, S. Deser and B. Zumino, ``{\sl Absence
of 3-Cocycles in the Dirac Monopole Problem}'', Phys. Lett. {\bf 153B}
(1985) 307;\\ Bo-Yu Hou, Bo-Yuan Hou and Pei Wang, ``{\sl How to
Eliminate the Dilemma in 3-Cocycle}'', Ann. Phys.
 (N.Y.) {\bf 171} (1986) 172.
%
\bibitem{chiao}
A. Tomita and R. Y. Chiao, ``{\sl Observation of Berry's Topological Phase by
Use of an Optical Fiber}'', Phys. Rev. Lett. {\bf 57} (1986) 937.
%
\bibitem{craige} ``{\sl Monopoles in Quantum Field Theory}'', Eds. N. S.
Craige, P. Goddard and W. Nahm (World Scientific, 1982).
%
\bibitem{mandel1} L. Mandel, ``{\sl Configuration-Space Photon Number
Operators in Quantum Optics}'', Phys. Rev. {\bf 144} (1966) 1071 and
``{\sl Photon Interference and Correlation Effects Produced by Independent
Quantum Sources}'', Phys. Rev. {\bf A28} (1983) 929.
%
\bibitem{mandel2} C. K. Hong and L. Mandel, ``{\sl
Experimental Realization of a Localized One-Photon State}'', Phys. Rev.
Lett. {\bf 56} (1986) 58.
 \end{thebibliography}
 \end{document}